\documentclass[conference,10pt]{IEEEtran}
\IEEEoverridecommandlockouts
\def\BibTeX{{\rm B\kern-.05em{\sc i\kern-.025em b}\kern-.08em
    T\kern-.1667em\lower.7ex\hbox{E}\kern-.125emX}}

\usepackage[utf8]{inputenc}
\usepackage{cite}
\usepackage{amsmath,amssymb,amsfonts}
\usepackage{algorithmic}
\usepackage{graphicx}
\usepackage{textcomp}
\usepackage[caption=false]{subfig}
\usepackage[ruled,vlined]{algorithm2e}
\usepackage{epsfig}
\usepackage{float}
\usepackage{color}
\usepackage{balance}
\usepackage{bbm}
\usepackage{textcomp}
\usepackage{enumitem}
\usepackage{comment}

\usepackage{mathtools,nccmath,amsthm}
\usepackage{multicol,tikz,array,booktabs,url}
\usepackage{xcolor}
\usepackage{acronym}
\usepackage{xfrac}

\graphicspath{{figures/}{./}}
\DeclareGraphicsExtensions{.pdf}

\newcommand{\vect}[1]{\boldsymbol{\mathbf{#1}}}

\newtheorem{Thm}{Theorem}
\newtheorem{Cor}{Corollary}
\newtheorem{Lem}{Lemma}
\newtheorem{Prop}{Proposition}

\DeclareMathOperator{\diag}{diag}

\DeclareMathOperator{\trace}{Tr}
\DeclarePairedDelimiter{\norm}{\lVert}{\rVert}

\DeclareMathAlphabet{\mathpzc}{OT1}{pzc}{m}{it}

\begin{document}
\title{The Interference Broadcast Channel with Reconfigurable Intelligent Surfaces:\\ A Cooperative Sum-Rate Maximization Approach}

\author{Konstantinos D. Katsanos$^1$, Paolo Di Lorenzo$^2$, and George C. Alexandropoulos$^{1,3}$\\
$^1$Department of Informatics and Telecommunications, National and Kapodistrian University of Athens, Greece
\\$^2$Department of Information Engineering, Electronics, and Telecommunications, Sapienza University, Italy
\\$^3$Department of Electrical and Computer Engineering, University of Illinois Chicago, IL, USA
\\e-mails: \{kkatsan, alexandg\}@di.uoa.gr, paolo.dilorenzo@uniroma1.it 
\vspace{-0.45cm}
\thanks{This work has been supported by the SNS JU project 6G-DISAC under the EU's Horizon Europe research and innovation program under Grant Agreement No 101139130.} \vspace{-0.45cm}
}

\maketitle
\thispagestyle{empty}

\begin{abstract}
This paper studies the interference broadcast channel comprising multiple multi-antenna Base Stations (BSs), each controlling a beyond diagonal Reconfigurable Intelligent Surface (RIS) and serving multiple single-antenna users. Wideband transmissions are considered with the objective to jointly design the BS linear precoding vectors and the phase configurations at the RISs in a distributed manner. We take into account the frequency selectivity behavior of each RIS's tunable meta-element, and focusing on the sum rate as the system's performance criterion, we present a distributed optimization approach that enables cooperation between the RIS control units and their respective BSs. According to the proposed scheme, each design variable can be efficiently obtained in an iterative parallel way with guaranteed convergence properties. Our simulation results demonstrate the validity of the presented distributed algorithm and showcase its superiority over a non-cooperative scheme as well as over the special case where the RISs have a conventional diagonal structure.
\end{abstract}

\vspace{-0.30cm}
\section{Introduction} \label{Sec:Intro}\vspace{-0.15cm}
Future generation wireless networks are required to reliably support massive connectivity and meet ultra-high data rate demands in a multi-functional intelligent manner~\cite{CMY+24}. As a result, efficient and low-cost physical-layer technologies, capable to overcome the existing algorithmic and infrastructure limitations, are of paramount importance. Lately, the technology of Reconfigurable Intelligent Surfaces (RISs)~\cite{RISoverview2023} has attracted considerable research interest as a strong candidate technology for enhancing coverage extension and achieving significant improvements for various other design objectives~\cite{RIS_challenges}. Very recently, the concept of Beyond Diagonal (BD) RISs has emerged \cite{LSN+23}, as a means to further increase the technology's achievable performance gains, providing more degrees of freedom for impacting over-the-air signal programmability.

Orthogonal Frequency Division Multiplexing (OFDM) constitutes the standardized wideband transmission method for current wireless communication systems, whose consideration in conjunction with RISs constitutes a direction worth studying~\cite{katsanos2022wideband}. To this end, very recently, the frequency-dependent behavior of BD RISs was investigated in~\cite{DRM+24,li2024wideband}, focusing on a multi-band multi-Base Station (BS) single BD-RIS-assisted network and a single-user system, respectively. Both studies advocated on the necessity to consider the frequency selectivity profile of each metamaterial element in practice.

In this paper, we focus on the sum-rate maximization design of an interference broadcast channel consisting of multi-antenna BSs and multiple BD RISs. Taking into account the frequency selectivity property of each RIS element, we devise a parallel cooperative design of the linear precoding vectors at the BSs and the phase configurations at the BD metasurfaces. Our numerical investigations demonstrate that the proposed distributed design leads to superior performance with respect to non-cooperative schemes and the case of diagonal RISs. 

\textit{Notations:} Boldface lower-case and upper-case letters represent vectors and matrices, respectively. The transpose, Hermitian transpose, conjugate, and the real part of a complex quantity are represented by $(\cdot)^{\rm T}$,  $(\cdot)^{\rm H}$, $(\cdot)^*$, and $\Re\{ \cdot \}$, respectively, while $\mathbb{C}$ is the set of complex numbers, and $\jmath\triangleq\sqrt{-1}$ is the imaginary unit. The symbols $<\!\cdot,\cdot>$ and $\mathbb{E}\{\cdot\}$ denote the inner product, and the statistical expectation, respectively, and $\mathbf{x}\sim\mathcal{CN}(\mathbf{a},\mathbf{A})$ indicates a complex Gaussian random vector with mean $\mathbf{a}$ and covariance matrix $\mathbf{A}$. $\operatorname{diag}\{\mathbf{a}\}$ is defined as the matrix whose diagonal elements are the entries of $\mathbf{a}$, while $\operatorname{vec}_{\rm d}(\vect{A})$ denotes the vector obtained by the diagonal elements of the square matrix $\vect{A}$. $\nabla_{\mathbf{a}}f$ denotes the Euclidean gradient vector of function $f$ along the direction indicated by $\mathbf{a}$.\vspace{-0.15cm}


\section{System Model and Problem Formulation} \label{Sec:Sys_Model_and_Problem}\vspace{-0.15cm}
\subsection{System Model} \label{Sec:Sys_Model}\vspace{-0.15cm}
We consider a multi-RIS-empowered interference broadcast channel comprising $Q$ multi-antenna BSs, each wishing to communicate in the downlink direction with multiple single-antenna User Equipments (UEs). We assume that each $N$-antenna BS sends information to its exclusively associated UEs using OFDM in a common set of physical resources, e.g., time and bandwidth. Thus, each BS-UE communicating pair is modeled as the superposition of a direct BS-UE link and a BS-RIS-UE link realized via the RIS-enabled tunable reflection. Each RIS, comprising $M$ passive reflecting elements, is assumed to be controlled by its solely owned BS and is placed either closely to it or near to the corresponding set of UEs~\cite{RIS_challenges}.

According to the deployed OFDM scheme, the total bandwidth is equally split into $K$ orthogonal Sub-Carriers (SCs). Let $\vect{w}_{\ell_q,k} \in \mathbb{C}^{N \times 1}$, with $k=1,2,\ldots,K$ represent the linear precoding vector at each $q$-th BS that models the digital spatial processing of its unit-power signal $s_{\ell_q,k}$ (i.e., $\mathbb{E}\{|s_{\ell_q,k}|^2\} = 1$) before transmission. We assume that the total transmit power available at each $q$-th BS is given by $P_q$. Letting $L_q$ denote the number of assigned UEs to the $q$-th BS, the corresponding transmit signal $\vect{x}_{q,k}$ can be compactly expressed as: $\vect{x}_{q,k} = \sum_{\ell=1}^{L_q} \vect{w}_{\ell_q,k} s_{\ell_q,k}$. Thus, the condition $\sum_{\ell=1}^{L_q}\sum_{k=1}^K \norm{\vect{w}_{\ell_q,k}}^2 \leq P_q$ must be satisfied. We also consider a quasi-static block fading channel model for all channels involved and focus on each particular fading block where the channels remain approximately constant with perfect Channel State Information (CSI) knowledge.\vspace{-0.08cm}

\subsection{BD RIS Structure and Element Response} \label{Sec:Freq_Response}\vspace{-0.10cm}
We consider a BD RIS structure~\cite{li2022_nonDiag_switches}, according to which an $M\times M$ array of ON/OFF-state switches is deployed to interconnect all RIS elements. Specifically, an ON-state at the switch in the position $(i,j)$ ($i,j=1,2,\ldots,M$) of the switch array indicates that the signal impinging on the $i$-th metamaterial element will be guided to and tunably reflected by the $j$-th element. This behavior can be mathematically expressed by a selection matrix $\vect{S}_q \in \{0,1\}^{M\times M}$ ($q=1,2,\ldots,Q$), whose role is to indicate the switch array selection process at each $q$-th RIS. In particular, each $\vect{S}_q$ is a binary-valued selection matrix (i.e., $[\vect{S}_q]_{i,j} \in \{0,1\}$) which by definition needs to satisfy the property of having only one non-zero value per row and column simultaneously and, thus, constitutes an extra design parameter. Clearly, a typical diagonal RIS, which does not require switches~\cite{RISoverview2023}, is obtained by setting $\vect{S}_q=\mathbf{I}_M$.

We further make the quite general assumption that each $m$-th unit element ($m=1,2,\ldots,M$) of each $q$-th RIS can be characterized as an equivalent parallel resonant circuit comprising a resistor $R$, a tunable capacitor $C_{mq}$, and two inductors $\mathpzc{L}_1$ and $\mathpzc{L}_2$ \cite{abeywickrama2020intelligent}. Then, the response of each $m$-th unit element of each $q$-th RIS is given by the reflection coefficient which is expressed in the frequency domain as follows:
\begin{equation} \label{eqn:reflect_coeff}
	\phi_{mq}(f,C_{mq}) = \frac{\mathcal{Z}(f,C_{mq}) - \mathcal{Z}_0}{\mathcal{Z}(f,C_{mq}) + \mathcal{Z}_0},
\end{equation}
where $\mathcal{Z}_0$ is the free space impedance, while $\mathcal{Z}(f,C_{mq})$ denotes the characteristic impedance of the equivalent circuit which is given for $\kappa\triangleq2\pi$ by
\begin{equation} \label{eqn:characteristic_impedance}
	\mathcal{Z}(f,C_{mq}) = \frac{\jmath \kappa f \mathpzc{L}_1\left(\jmath \kappa f \mathpzc{L}_2 + R + \frac{1}{\jmath \kappa f C_{mq}}\right)}{\jmath \kappa f \left(\mathpzc{L}_1+\mathpzc{L}_2\right) + R + \frac{1}{\jmath \kappa f C_{mq}}}.
\end{equation}
Instead of constructing fitting functions \cite{li2024wideband} to simplify the manipulations of the latter highly non-linear function with respect to the tunable parameters $C_{mq}$, in this work, we observe that \eqref{eqn:reflect_coeff} can be equivalently transformed into a more tractable form, according to the following proposition.
\begin{Prop} \label{prop:RIS_Freq_Response}
	The frequency response of each $m$-th unit element of each $q$-th RIS can be reformulated as follows:
	\begin{equation} \label{eqn:RIS_freq_response}
		\phi_{mq}(f,C_{mq}) = 1 - \frac{2}{1 + \frac{\mathcal{D}_{mq}(f,C_{mq})}{\mathcal{N}_{mq}(f,C_{mq})}},
	\end{equation}
	where $\mathcal{N}_{mq}(f,C_{mq})$ and $\mathcal{D}_{mq}(f,C_{mq})$ are defined as follows:
	\begin{align}
		\mathcal{N}_{mq}(f,C_{mq}) \!&\triangleq\! 1 \!-\! (\kappa f)^2(\mathpzc{L}_1 \!+\! \mathpzc{L}_2)C_{mq} \!+\! \jmath\kappa f R C_{mq}, \label{eqn:numerator_s} \\ 
		\mathcal{D}_{mq}(f,C_{mq}) \!&\triangleq\! \jmath \kappa f\frac{\mathpzc{L}_1}{\mathcal{Z}_0}\left(1\!-\!(\kappa f)^2 \mathpzc{L}_2 C_{mq}\!+\!\jmath \kappa f R C_{mq} \right)\!. \label{eqn:denominator_s}
	\end{align}
\end{Prop}
\begin{IEEEproof}
	The expression \eqref{eqn:RIS_freq_response} follows by replacing \eqref{eqn:characteristic_impedance} into \eqref{eqn:reflect_coeff} and straightforward algebraic manipulations.
\end{IEEEproof} \vspace{-0.15cm}

\subsection{Received Signal Model} \label{Sec:Received_Model}\vspace{-0.15cm}
Based on the considered system model, for each $\ell_q$-th BS-UE pair, there will be an additional RIS-enabled wireless link, through which the signals transmitted by the $q$-th BS are reflected by its owned $q$-th RIS before arriving at the intended $\ell_q$-th UE. Let $\vect{H}_{q,q,k} \in \mathbb{C}^{M\times N}$ and $\vect{g}_{q,\ell_q,k} \in \mathbb{C}^{M \times 1}$ denote each $q$-th BS-RIS and each $\ell_q$-th RIS-UE channel, respectively, at each $k$-th SC. We define the vector $\vect{\phi}_{q,k} \triangleq [\phi_{1q}(f_k,C_{1q}),\ldots,\phi_{Mq}(f_k,C_{Mq})]^{\rm T} \in \mathbb{C}^{M \times 1}$ as the one including the reflection coefficients of each $q$-th RIS and then formulate the matrices $\vect{\Phi}_{q,k} \triangleq \operatorname{diag}\{\vect{\phi}_{q,k}\}$ $\forall$$q,k$. Then, the baseband received signal at each $\ell_q$-th UE at each $k$-th SC (i.e., in the frequency domain) can be expressed as:
\begin{equation} \label{eqn:received_signal}
	y_{\ell_q,k} = \vect{f}_{q,\ell_q,k}^{\rm H} \vect{x}_{q,k} + \sum_{j\neq q}^Q \vect{f}_{j,\ell_q,k}^{\rm H} \vect{x}_{j,k} + n_{\ell_q,k},
\end{equation}
where $n_{\ell_q,k} \sim \mathcal{CN}(0,\sigma_{\ell_q,k}^2)$ represents the Additive White Gaussian Noise (AWGN), which models the thermal noises at the UE receivers. We have also used the definitions:
\begin{align} 
	\vect{f}_{q,\ell_q,k}^{\rm H} &\triangleq \vect{h}_{q,\ell_q,k}^{\rm H} + \vect{g}_{q,\ell_q,k}^{\rm H}\vect{S}_q\vect{\Phi}_{q,k}\vect{H}_{q,q,k}, \label{eqn:total_channels_1}\\
	\vect{f}_{j,\ell_q,k}^{\rm H} &\triangleq \vect{h}_{j,\ell_q,k}^{\rm H} + \vect{g}_{j,\ell_q,k}^{\rm H}\vect{S}_j\vect{\Phi}_{j,k}\vect{H}_{j,j,k},
	\label{eqn:total_channels_2}
\end{align}
where each $\vect{h}_{j,\ell_q,k}\in \mathbb{C}^{N \times 1}$ indicates the direct channel between the $\ell_q$-th UE and the $j$-th BS at each $k$-th SC. \vspace{-0.30cm}


\subsection{Problem Formulation} \label{Sec:Prob_Form}\vspace{-0.15cm}
We, first, define the vectors: \textit{i}) $\widetilde{\vect{w}} \triangleq [\tilde{\vect{w}}_1^{\rm T},\dots,\tilde{\vect{w}}_Q^{\rm T}]^{\rm T}$ with $\tilde{\vect{w}}_q \triangleq [\vect{w}_{1}^{\rm T},\dots,\vect{w}_{L_q}^{\rm T}]^{\rm T}$ and $\vect{w}_{\ell_q} \triangleq [\vect{w}_{\ell_q,1}^{\rm T},\ldots,\vect{w}_{\ell_q,K}^{\rm T}]^{\rm T}$; and \textit{ii}) $\widetilde{\vect{c}}\triangleq [\vect{c}_1^{\rm T},\ldots,\vect{c}_Q^{\rm T}]^{\rm T}$ with $\vect{c}_q \triangleq [C_{1q},\ldots,C_{Mq}]^{\rm T} \in \mathbb{R}^{M \times 1}$; as well as \textit{iii}) the set of matrices $\widetilde{\vect{S}} \triangleq \{\vect{S}_q\}_{q=1}^Q$ including, respectively, the precoding vectors at the $Q$ multi-antenna BSs, the tunable capacitances, and the switch selection matrices at the $Q$ BD RISs. Then, treating the Multi-User Interference (MUI) term in \eqref{eqn:received_signal} as an additional source of noise (colored noise), the achievable sum-rate performance in bits per second per Hertz (bits/s/Hz) for each $\ell_q$-th UE can be expressed as the following function of the tunable parameter triplet $(\widetilde{\vect{w}},\widetilde{\vect{c}},\widetilde{\vect{S}})$: 
\begin{equation} \label{eqn:sum_rate_q}
	\mathcal{R}_{\ell_q}\left(\widetilde{\vect{w}},\widetilde{\vect{c}},\widetilde{\vect{S}}\right) = \frac{1}{K}\sum_{k=1}^K \log_2\left( 1 + \frac{|\vect{f}_{q,\ell_q,k}^{\rm H}\vect{w}_{\ell_q,k}|^2}{\operatorname{MUI}_{\ell_q,k}} \right),
\end{equation}
with $\operatorname{MUI}_{\ell_q,k} \triangleq \sigma_{\ell_q,k}^2 + \sum_{(n,j)\neq(\ell,q)} |\vect{f}_{j,\ell_q,k}^{\rm H}\vect{w}_{n_j,k}|^2$, where the summation term can be decomposed as:
\begin{equation} \label{eqn:intra_intercell_interference}
	\begin{aligned}
		\underbrace{\sum_{m=1,m\neq\ell}^{L_q} |\vect{f}_{q,\ell_q,k}^{\rm H}\vect{w}_{m_q,k}|^2}_{\text{intracell interference}} + \underbrace{\sum_{j\neq q}^Q \sum_{n=1}^{L_j} |\vect{f}_{j,\ell_q,k}^{\rm H}\vect{w}_{n_j,k}|^2}_{\text{intercell interference}}.
	\end{aligned}	
\end{equation}
The dependence on $\widetilde{\vect{c}}$ and $\widetilde{\vect{S}}$ is implied via the composite channels $\vect{f}_{q,\ell_q,k}$ and $\vect{f}_{j,\ell_q,k}$ in \eqref{eqn:total_channels_1} and \eqref{eqn:total_channels_2}, respectively.

In this paper, we aim to maximize the achievable sum-rate performance of the proposed multi-RIS-empowered wireless system and consider the following optimization problem:
\begin{align*}
	\mathcal{OP}: \,\max_{\widetilde{\vect{w}},\widetilde{\vect{c}},\widetilde{\vect{S}}} \, & \quad \sum_{q=1}^Q \sum_{\ell=1}^{L_q} \mathcal{R}_{\ell_q}\left(\widetilde{\vect{w}},\widetilde{\vect{c}},\widetilde{\vect{S}}\right) \\
	\text{s.t.} & \quad \sum_{\ell=1}^{L_q}\sum_{k=1}^{K} \norm{\vect{w}_{\ell_q,k}}^2 \leq P_q, \vect{S}_q \in \mathcal{S},\forall q = 1,\ldots,Q,\\
	& \quad C_{\min} \leq [\vect{c}_q]_m \leq C_{\max},  \;\; \forall m=1,\ldots,M,
\end{align*}
where $\mathcal{S} \triangleq\left\{ \vect{S}\in \{0,1\}^{M\times M}:\vect{S}\vect{1} = \vect{1}, \vect{S}^{\rm T}\vect{1}=\vect{1} \right\}$ indicates the feasible set for the switch selection matrices at the BD RISs, while $C_{\min}$ and $C_{\max}$ represent the minimum and maximum allowable values for the RIS tunable capacitances according to circuital characteristics, respectively. 


\section{Distributed Sum-Rate Maximization} \label{Sec:Design_SCA}
Let $\vect{X}_q \triangleq \{\tilde{\vect{w}}_q,\vect{c}_q,\vect{S}_q\}$ and $\vect{X}_{-q}$ be the set of all other variables except the $q$-th triplet. The objective function in $\mathcal{OP}$ is non-concave, due to the presence of MUI and the coupling between the design variables. Nevertheless, we note that the sum-rate objective in $\mathcal{OP}$ can be decomposed into the following form:
\begin{equation} \label{eqn:total_rate}
	\overline{\mathcal{R}}(\vect{X}_q,\vect{X}_{-q})\!\triangleq\!\sum_{\ell=1}^{L_q}\mathcal{R}_{\ell_q}(\vect{X}_q,\vect{X}_{-q})+\sum_{j\neq q}^Q\sum_{\ell=1}^{L_j} \mathcal{R}_{\ell_j}(\vect{X}_q,\vect{X}_{-q}).
\end{equation}
The above structure leads to the following decomposition scheme similar to \cite{scutari2013decomposition}: \textit{i}) at every iteration $t$, the first set of terms (equal to $\mathcal{R}_q(\vect{X}_q,\vect{X}_{-q})$) is replaced by a surrogate function, denoted as $\widetilde{\mathcal{R}}_q(\vect{X}_q,\vect{X}^t)$, which depends on the current iterate $\vect{X}^t$; and \textit{ii}) the remaining terms involved are linearized around $\vect{X}_q^t$. Thus, the proposed updating scheme for distributedly solving $\mathcal{OP}$ reads as: at each algorithmic iteration $t$, each BS solves the optimization problem below:
\begin{equation*}\label{eqn:Surrogate_problem}
	\mathcal{OP}_1:\quad\widehat{\vect{X}}_q^t\,=\, \arg\max_{\vect{X}_q \in \mathcal{X}_q} \, \widetilde{\mathcal{R}}_q(\vect{X}_q;\vect{X}^t)+<\boldsymbol{\Pi}^t_q,\vect{X}_q -\vect{X}_q^t>, 
\end{equation*}
where $\mathcal{X}_q$ denotes the feasible set combining all constraints of $\mathcal{OP}$, while the local surrogate function $\widetilde{\mathcal{R}}_q$ is given by:
\begin{align} \label{eqn:first_surrogate}
	\begin{split}		
		&\widetilde{\mathcal{R}}_q(\vect{X}_q;\vect{X}^t) \triangleq \sum_{\ell=1}^{L_q}\sum_{k=1}^K \log_2 \left( 1 + \frac{|\vect{f}_{q,\ell_q,k}^{\rm H}\vect{w}_{\ell_q,k}|^2}{\operatorname{MUI}_{\ell_q,k}^t} \right) \\
		&+<\vect{\gamma}_{\vect{c}_q}^t,\vect{c}_q - \vect{c}_q^t> + <\vect{\Gamma}_{\vect{S}_q}^t,\vect{S}_q - \vect{S}_q^t> \\
		&-\frac{\tau}{2}\Big(\norm{\vect{w}_{\ell_q} - \vect{w}_{\ell_q}^t }^2+ \norm{\vect{c}_q - \vect{c}_q^t}^2 + \norm{ \vect{S}_q - \vect{S}_q^t}_{\rm F}^2\Big),
	\end{split}
\end{align}
with $\tau\!>\!0$ being an appropriately chosen parameter, $\vect{\gamma}_{\vect{c}_q}^t \triangleq  \nabla_{\vect{c}_q}\mathcal{R}_q(\vect{X}_q,\vect{X}_{-q}^t)\vert_{\vect{c}_q=\vect{c}_q^t}$ and accordingly for $\vect{\Gamma}_{\vect{S}_q}^t$. In addition, $\vect{\Pi}_q^t \triangleq \sum_{j\neq q}^Q \sum_{\ell=1}^{L_j} \nabla_{\vect{X}_q} \mathcal{R}_{\ell_j}(\vect{X}_q,\vect{X}_{-q})$ evaluated at $\vect{X}_q^t$, which is often referred to as the pricing vector/matrix \cite{scutari2013decomposition}. The multiplicative factor $1/K$ is ignored, since it does not affect the optimization solution approach. Next, we solve $\mathcal{OP}_1$ for each set of variables included in $\vect{X}_q$. 

\subsection{Local Linear Precoding Optimization} \label{Sec:Precoder}\vspace{-0.15cm}
Solving $\mathcal{OP}_1$ with respect to the linear precoder $\vect{w}_{\ell_q}$ for the $\ell_q$-th UE leads to the following optimization sub-problem:
\begin{align*}
	\begin{split}
		\mathcal{OP}_{\vect{w}_{\ell_q}}: \,\max_{\vect{w}_{\ell_q}} \, & \,\, \sum_{\ell=1}^{L_q}\Bigg(\sum_{k=1}^K \breve{\mathcal{R}}_{\ell_q,k}(\vect{w}_{\ell_q,k}) - \frac{\tau}{2}\norm{\vect{w}_{\ell_q} - \vect{w}_{\ell_q}^t}^2 \\
		&+ \Re \left\{ (\overline{\vect{\pi}}_{\ell_q}^t)^{\rm H}(\vect{w}_{\ell_q} - \vect{w}_{\ell_q}^t) \right\}\Bigg)\\
		\text{s.t.} & \,\, \sum_{\ell=1}^{L_q}\sum_{k=1}^K \norm{\vect{w}_{\ell_q,k}}^2 \leq P_q,
	\end{split}
\end{align*}
where $\breve{\mathcal{R}}_{\ell_q,k}(\vect{w}_{\ell_q,k})$ stands for the logarithmic term in \eqref{eqn:first_surrogate}. Also, $\overline{\vect{\pi}}_{\ell_q}^t$ is the pricing vector associated with $\vect{w}_{\ell_q}$, which is given by $\overline{\vect{\pi}}_{\ell_q}^t = [(\overline{\vect{\pi}}_{\ell_q,1}^t)^T,\ldots,(\overline{\vect{\pi}}_{\ell_q,K}^t)^T]^T$, with
\begin{equation} \label{eqn:pricing_w_q}
		\overline{\vect{\pi}}_{\ell_q,k}^t \!=\! \sum_{j \neq q}^Q \sum_{n=1}^{L_j} \frac{-\operatorname{snr}_{n_j,k}^t/\ln(2)}{(1 + \operatorname{snr}_{n_j,k}^t)\operatorname{MUI}_{n_j,k}^t}\vect{f}_{q,n_j,k} \vect{f}_{q,n_j,k}^{\rm H} \vect{w}_{\ell_q,k}^t,
\end{equation}
where $\operatorname{snr}_{n_j,k}^t \triangleq \left\lvert \vect{f}_{j,n_j,k}^{\rm H}\vect{w}_{n_j,k}^t \right\rvert^2/\operatorname{MUI}_{n_j,k}^t$.

$\mathcal{OP}_{\vect{w}_{\ell_q}}$ is still a non-concave problem, primarily owing to the logarithmic function involving the quadratic term with respect to $\vect{w}_{\ell_q,k}$ in $\breve{\mathcal{R}}_{\ell_q,k}$. To address this challenge and maintain a valid surrogate function that retains first-order properties, we make use of the following Lemma.
\begin{Lem} \label{lem:logarithmic_surrogate}
	The logarithmic term $\breve{\mathcal{R}}_{\ell_q,k}(\vect{w}_{\ell_q,k})$ in \eqref{eqn:first_surrogate} can be lower-bounded by the following surrogate function: 
	\begin{equation} \label{eqn:logarithmic_surrogate}
		\begin{aligned}
			\widehat{\mathcal{R}}_{\ell_q,k} \!=\! -a_{\ell_q,k}^t\vect{w}_{\ell_q,k}^{\rm H} \vect{F}_{\ell_q,k} \vect{w}_{\ell_q,k} + 2\Re\{(\vect{b}_{\ell_q,k}^t)^{\rm H}\vect{w}_{\ell_q,k}\},
		\end{aligned}
	\end{equation}
	where $\vect{F}_{\ell_q,k} \triangleq \vect{f}_{q,\ell_q,k}\vect{f}_{q,\ell_q,k}^{\rm H}$, $a_{\ell_q,k}^t$ and $\vect{b}_{\ell_q,k}^t$ are defined as:
	\begin{align}
		a_{\ell_q,k}^t &\triangleq \frac{1}{\ln(2)}\frac{\lvert \vect{f}_{q,\ell_q,k}^{\rm H} \vect{w}_{\ell_q,k}^t \rvert^2}{(\operatorname{MUI}_{\ell_q,k}^t + \lvert \vect{f}_{q,\ell_q,k}^{\rm H} \vect{w}_{\ell_q,k}^t \rvert^2) \operatorname{MUI}_{\ell_q,k}^t}, \label{eqn:surrog_a} \\
		\vect{b}_{\ell_q,k}^t &\triangleq \frac{1}{\ln(2)}\frac{1}{\operatorname{MUI}_{\ell_q,k}^t} \vect{F}_{\ell_q,k} \vect{w}_{\ell_q,k}^t. \label{eqn:surrog_b}
	\end{align}
\end{Lem}
\begin{IEEEproof}
	The proof follows from the observation that $\breve{\mathcal{R}}_{\ell_q,k}=-\log_2(1 - c_{\ell_q,k}^{-1}|d_{\ell_q,k}|^2)$, where $c_{\ell_q,k}\triangleq |d_{\ell_q,k}|^2 + \operatorname{MUI}_{\ell_q,k}^t$ and $d_{\ell_q,k} \triangleq \vect{f}_{q,\ell_q,k}^H\vect{w}_{\ell_q,k}$. Then, the modified logarithmic function is a jointly convex function with respect to $(c_{\ell_q,k},d_{\ell_q,k})$ \cite{tam2016_successive}. Thus, it suffices to derive the first-order Taylor expansion around the feasible point $(c_{\ell_q,k}^t,d_{\ell_q,k}^t)$.
\end{IEEEproof}

Next, exploiting Lemma~\ref{lem:logarithmic_surrogate} and defining the block diagonal matrix $\tilde{\vect{F}}_{\ell_q} \triangleq \operatorname{blkdiag}\{a_{\ell_q,k}^t\vect{F}_{\ell_q,k}\}_{k=1}^K$, and the vector $\tilde{\vect{f}}_{\ell_q} \triangleq [(\vect{b}_{\ell_q,1}^t)^{\rm T},\ldots,(\vect{b}_{\ell_q,K}^t)^{\rm T}]^{\rm T}$, the objective in $\mathcal{OP}_{\vect{w}_{\ell_q}}$ becomes:
\begin{equation} \label{eqn:compact_surrogate_w}
	\mathcal{J}=-\vect{w}_{\ell_q}^{\rm H}\left(\tilde{\vect{F}}_{\ell_q} + \frac{\tau}{2}\vect{I}_{KN} \right)\vect{w}_{\ell_q} + \Re\left\{(\vect{v}_{\ell_q}^t)^{\rm H}\vect{w}_{\ell_q} \right\},
\end{equation}
where $\vect{v}_{\ell_q}^t \triangleq \overline{\vect{\pi}}_{\ell_q}^t + 2\tilde{\vect{f}}_{\ell_q} + \tau \vect{w}_{\ell_q}^t$. It can be deduced that $\tilde{\vect{F}}_{\ell_q}\succeq\vect{0}$ yielding the concavity of \eqref{eqn:compact_surrogate_w}. Therefore, the optimal $\vect{w}_{\ell_q}$ follows by the first-order condition, which results in:
\begin{equation} \label{eqn:optimal_w}
	\vect{w}_{\ell_q}^{\rm opt}(\lambda) = \left(\tilde{\vect{F}}_{\ell_q} + \left(\frac{\tau}{2} + \lambda\right)\vect{I}_{KN} \right)^{-1} \vect{v}_{\ell_q}^t,
\end{equation}
where $\lambda \geq 0$ denotes the Lagrange multiplier associated with the transmit power constraint, whose optimum value ($\lambda^{\rm opt}$) can be obtained by elaborating on Slater's condition and a bisection search, similarly to \cite[Corollary 1]{PLS2022_counteracting}. \vspace{-0.60cm}

\subsection{Local RIS Reflection Configuration Optimization} \label{Sec:RIS_Config}\vspace{-0.25cm}
The reflection configuration vector at each $q$-th RIS for each $k$-th SC (i.e., $\vect{\phi}_{q,k}$) is a function of the parameters included in $\vect{c}_q$, and can be optimized by solving the following problem:
\begin{align*}
	\mathcal{OP}_{\vect{c}_q}: \,\max_{\vect{c}_q} \, & \,\, - \frac{\tau}{2}\norm{\vect{c}_q - \vect{c}_q^t}^2 + \Re\{ (\vect{\gamma}_{\vect{c}_q}^t + \underline{\vect{\pi}}_q^t)^{\rm H}(\vect{c}_q - \vect{c}_q^t)\} \\
	\text{s.t.} & \quad C_{\min} \leq [\vect{c}_q]_m \leq C_{\max} \, \, \forall m=1,2,\ldots,M,
\end{align*}
which is clearly a concave optimization problem. Before proceeding to its solution, the analytic expressions for $\vect{\gamma}_{\vect{c}_q}^t$ and $\underline{\vect{\pi}}_q^t$ are presented in the next Theorem.

\begin{Thm} \label{thm:RIS_vectors} \vspace{-0.15cm}
	Let the following matrix definitions:\vspace{-0.30cm}
	\begin{align}
		\vect{A}_{q,\ell_q,k} &\triangleq \vect{H}_{q,q,k}\vect{w}_{\ell_q,k}\vect{w}_{\ell_q,k}^{\rm H}\vect{h}_{q,\ell_q,k}\vect{g}_{q,\ell_q,k}^{\rm H}\vect{S}_q, \\
		\vect{A}_{q,\ell_q,k}^{m_q} &\triangleq \vect{H}_{q,q,k}\vect{w}_{m_q,k}\vect{w}_{m_q,k}^{\rm H}\vect{h}_{q,\ell_q,k}\vect{g}_{q,\ell_q,k}^{\rm H}\vect{S}_q, \\
		\vect{A}_{q,n_j,k} &\triangleq \vect{H}_{q,q,k}\vect{w}_{\ell_q,k}\vect{w}_{\ell_q,k}^{\rm H}\vect{h}_{q,n_j,k}\vect{g}_{q,n_j,k}^{\rm H}\vect{S}_q, \\
		\vect{B}_{q,\ell_q,k} &\triangleq \vect{S}_q^{\rm T}\vect{g}_{q,\ell_q,k}\vect{g}_{q,\ell_q,k}^{\rm H}\vect{S}_q, \\
		\vect{B}_{q,n_j,k} &\triangleq \vect{S}_q^{\rm T}\vect{g}_{q,n_j,k}\vect{g}_{q,n_j,k}^{\rm H}\vect{S}_q, \\
		\vect{C}_{q,\ell_q,k} &\triangleq \vect{H}_{q,q,k}\vect{w}_{\ell_q,k}\vect{w}_{\ell_q,k}^{\rm H}\vect{H}_{q,q,k}^{\rm H}, \\
		\vect{C}_{q,m_q,k} &\triangleq \vect{H}_{q,q,k}\vect{w}_{m_q,k}\vect{w}_{m_q,k}^{\rm H}\vect{H}_{q,q,k}^{\rm H}, \\
		\vect{M}_{q,\ell_q,k} &\triangleq \vect{A}_{q,\ell_q,k} + \vect{C}_{q,\ell_q,k} (\vect{\Phi}_{q,k}^t)^{\rm H} \vect{B}_{q,\ell_q,k}, \\
        \vect{M}_{q,\ell_q,k}^{m_q} &\triangleq \vect{A}_{q,\ell_q,k}^{m_q} + \vect{C}_{q,m_q,k} (\vect{\Phi}_{q,k}^t)^{\rm H} \vect{B}_{q,\ell_q,k}, \\
		\vect{M}_{q,n_j,k} &\triangleq \vect{A}_{q,n_j,k} + \vect{C}_{q,\ell_q,k} (\vect{\Phi}_{q,k}^t)^{\rm H} \vect{B}_{q,n_j,k}, \\
		\vect{Q}_{q,k} &\triangleq \diag\left\{\frac{\partial([\vect{\phi}_{qk}]_1)^*}{\partial C_{1q}},\ldots,\frac{\partial([\vect{\phi}_{qk}]_M)^*}{\partial C_{Mq}}\right\},
	\end{align}
	where the partial derivatives of $[\vect{\phi}_{q,k}]_m$ $\forall$$q,k,m$ with respect to RIS tunable capacitance $C_{mq}$ can be computed as:
	\begin{align} \label{eqn:der_C_m}  
		\nonumber\frac{\partial([\vect{\phi}_{q,k}]_m)^*}{\partial C_{mq}} =&\frac{-2}{\left(\mathcal{N}_{mq}^*(f_k,C_{mq}) + \mathcal{D}_{mq}^*(f_k,C_{mq}) \right)^2}\\
		&\times\Bigg( \frac{\partial \mathcal{N}_{mq}^*(f_k,C_{mq})}{\partial C_{mq}} \mathcal{D}_{mq}^*(f_k,C_{mq})\\
		&\hspace{0.68cm}- \mathcal{N}_{mq}^*(f_k,C_{mq}) \frac{\partial \mathcal{D}_{mq}^*(f_k,C_{mq})}{\partial C_{mq}} \Bigg),\nonumber
	\end{align}
	where respectively following \eqref{eqn:numerator_s} and \eqref{eqn:denominator_s} holds that: 
	\begin{align}
		\frac{\partial \mathcal{N}_{mq}^*(f_k,C_{mq})}{\partial C_{mq}} &= -(\kappa f_k)^2(\mathpzc{L}_1 + \mathpzc{L}_2) - \jmath \kappa f_k R, \\
		\frac{\partial \mathcal{D}_{mq}^*(f_k,C_{mq})}{\partial C_{mq}} &= -\jmath \kappa f_k \frac{\mathpzc{L}_1}{\mathcal{Z}_0}(-(\kappa f_k)^2 \mathpzc{L}_2 - \jmath \kappa f_k R).
	\end{align}
	Then, the vectors $\vect{\gamma}_{\vect{c}_{q}}^t$ and $\underline{\vect{\pi}}_{\vect{c}_{q}}^t$ in $\mathcal{OP}_{\vect{c}_q}$ are given by the following analytic expressions:\vspace{-0.15cm}
	\begin{align}
		&\begin{aligned}
			&\vect{\gamma}_{\vect{c}_q}^t = \sum_{\ell=1}^{L_q}\sum_{k=1}^K \frac{2/\ln(2)}{(1+\operatorname{snr}_{\ell_q,k}^t)(\operatorname{MUI}_{\ell_q,k}^t)^2} \\
			&\times \Bigg( \Re\Bigg\{ \operatorname{MUI}_{\ell_q,k}^t \vect{Q}_{q,k} \operatorname{vec}_{\rm d}(\vect{M}_{q,\ell_q,k}) \Bigg\} \\
			&- |\vect{f}_{q,\ell_q,k}^{\rm H}\vect{w}_{\ell_q,k}|^2 \sum_{m\neq\ell}^{L_q}\Re\Bigg\{\vect{Q}_{q,k}\operatorname{vec}_{\rm d}(\vect{M}_{q,\ell_q,k}^{m_q}) \Bigg\} \Bigg),
		\end{aligned} \label{eqn:gamma_RIS}\\
		&\begin{aligned}
			\underline{\vect{\pi}}_{\vect{c}_{q}}^t =& \sum_{j \neq q}^Q \sum_{n=1}^{L_j} \sum_{k=1}^K\frac{-(2/\ln(2))\operatorname{snr}_{n_j,k}^t}{(1+\operatorname{snr}_{n_j,k}^t)\operatorname{MUI}_{n_j,k}^t} \\
			&\times \Re\Bigg\{\vect{Q}_{q,k}\operatorname{vec}_{\rm d}(\vect{M}_{q,n_j,k})\Bigg\}. \label{eqn:pricing_RIS}
		\end{aligned}
	\end{align}
\end{Thm}\vspace{-0.15cm}
\begin{IEEEproof}
	The proof is based on \cite[Theorem 1]{katsanos2024multi}.
\end{IEEEproof}

Next, $\mathcal{OP}_{\vect{c}_q}$ can be solved in closed form.
\begin{Cor} \label{cor:RIS_cl_form_solution}
	Let $\vect{\beta}_q \triangleq \tau\vect{c}_q^t + \vect{\gamma}_{\vect{c}_q}^t + \underline{\vect{\pi}}_{\vect{c}_{q}}^t$. Then, $\mathcal{OP}_{\vect{c}_q}$'s optimal solution is given in closed form as follows:
	\begin{equation} \label{eqn:RIS_solution}
		[\vect{c}_q]_m^{\rm opt} =  \begin{cases}
			C_{\min}, & \text{if}\,\,\frac{1}{\tau}[\vect{\beta}_q]_m < C_{\min} \\
			C_{\max}, & \text{if}\,\,\frac{1}{\tau}[\vect{\beta}_q]_m >C_{\max} \\
			\frac{1}{\tau}[\vect{\beta}_q]_m, & \text{otherwise}
		\end{cases}.
	\end{equation}
\end{Cor} \vspace{-0.15cm}
\begin{IEEEproof}
	Follows by expanding $\mathcal{OP}_{\vect{c}_q}$'s objective.
\end{IEEEproof}\vspace{-0.15cm}

\subsection{Local RIS Switch Selection Matrix Optimization}\label{Sec:Non_Diag_Design} \vspace{-0.15cm}
The design of the switch selection matrix $\vect{S}_q$ at each $q$-th BD RIS, reduces to the following simplified optimization problem, by noting that $\trace(\vect{S}_q\vect{S}_q^{\rm T}) = M$:
\begin{align*}
	\mathcal{OP}_{\vect{S}_q}: \max_{\vect{S}_q \in \mathcal{S}} \, & \,\, \trace\left( \Re\left\{\vect{\Gamma}_{\vect{S}_q}^t + \vect{\Pi}_{\vect{S}_q}^t + \tau\vect{S}_q^t\right\}^{\rm H}\vect{S}_q \right),
\end{align*}
whose solution depends on $\vect{\Gamma}_{\vect{S}_q}^t$ and $\vect{\Pi}_{\vect{S}_q}^t$ derived below.
\begin{Cor} \label{thm:Sel_Mat_vectors}
	Let the following matrix definitions:
	\begin{align} 
		\vect{F}_{q,\ell_q,k} &\triangleq \vect{\Phi}_{q,k}\vect{H}_{q,q,k}\vect{w}_{\ell_q,k}\vect{w}_{\ell_q,k}^{\rm H}\vect{h}_{q,\ell_q,k}\vect{g}_{q,\ell_q,k}^{\rm H}, \\
		\vect{F}_{q,\ell_q,k}^{m_q} &\triangleq \vect{\Phi}_{q,k}\vect{H}_{q,q,k}\vect{w}_{m_q,k}\vect{w}_{m_q,k}^{\rm H}\vect{h}_{q,\ell_q,k}\vect{g}_{q,\ell_q,k}^{\rm H}, \\
		\vect{F}_{q,n_j,k} &\triangleq \vect{\Phi}_{q,k}\vect{H}_{q,q,k}\vect{w}_{\ell_q,k}\vect{w}_{\ell_q,k}^{\rm H}\vect{h}_{q,n_j,k}\vect{g}_{q,n_j,k}^{\rm H},  \\
		\vect{K}_{q,\ell_q,k} &\triangleq \vect{\Phi}_{q,k}\vect{H}_{q,q,k}\vect{w}_{\ell_q,k}\vect{w}_{\ell_q,k}^{\rm H}\vect{H}_{q,q,k}^{\rm H}\vect{\Phi}_{q,k}^{\rm H},  \\
		\vect{K}_{q,\ell_q,k}^{m_q} &\triangleq \vect{\Phi}_{q,k}\vect{H}_{q,q,k}\vect{w}_{m_q,k}\vect{w}_{m_q,k}^{\rm H}\vect{H}_{q,q,k}^{\rm H}\vect{\Phi}_{q,k}^{\rm H},  \\
		\vect{G}_{q,\ell_q,k} &\triangleq \vect{g}_{q,\ell_q,k}\vect{g}_{q,\ell_q,k}^{\rm H}, \,\,\vect{G}_{q,n_j,k} \triangleq \vect{g}_{q,n_j,k}\vect{g}_{q,n_j,k}^{\rm H}, \\
		\vect{N}_{q,\ell_q,k} &\triangleq \vect{F}_{q,\ell_q,k} + \vect{K}_{q,\ell_q,k} (\vect{S}_q^t)^T \vect{G}_{q,\ell_q,k}, \\
		\vect{N}_{q,\ell_q,k}^{m_q} &\triangleq \vect{F}_{q,\ell_q,k}^{m_q} + \vect{K}_{q,\ell_q,k}^{m_q} (\vect{S}_q^t)^T \vect{G}_{q,\ell_q,k}, \\
		\vect{N}_{q,n_j,k} &\triangleq \vect{F}_{q,n_j,k} + \vect{K}_{q,\ell_q,k} (\vect{S}_q^t)^T \vect{G}_{q,n_j,k}.
	\end{align}
	Then, $\vect{\Gamma}_{\vect{S}_q}^t$ and $\vect{\Pi}_{\vect{S}_q}^t$ are given by:
	\begin{align}
		&\begin{aligned}
			&\vect{\Gamma}_{\vect{S}_q}^t = \sum_{\ell=1}^{L_q}\sum_{k=1}^K \frac{2/\ln(2)}{(1+\operatorname{snr}_{\ell_q,k}^t)(\operatorname{MUI}_{\ell_q,k}^t)^2}\\
            &\times \Bigg(\!\operatorname{MUI}_{\ell_q,k}^t\vect{N}_{q,\ell_q,k}-|\vect{f}_{q,\ell_q,k}^{\rm H}\vect{w}_{\ell_q,k}|^2\sum_{m\neq\ell}^{L_q}\vect{N}_{q,\ell_q,k}^{m_q}\!\Bigg)^{\rm T}, \\
		\end{aligned}  \label{eqn:gamma_Sel_Mat}\\
		&\begin{aligned} 
			\vect{\Pi}_{\vect{S}_q}^t = &\sum_{j\neq q}^Q \sum_{n=1}^{L_j} \sum_{k=1}^K\frac{-(2/\ln(2))\operatorname{snr}_{n_j,k}^t}{(1+\operatorname{snr}_{n_j,k}^t)\operatorname{MUI}_{n_j,k}^t} \vect{N}_{q,n_j,k}^{\rm T}.
		\end{aligned} \label{eqn:pricing_Sel_Mat}
	\end{align}
\end{Cor}
\begin{IEEEproof}
	The proof is based on \cite[Corollary 2]{katsanos2024multi}.
\end{IEEEproof}

$\mathcal{OP}_{\vect{S}_q}$ can be tackled, without loss of optimality, by dropping the binary constraints and relaxing the rest of them. Then, it can be efficiently solved as a linear program.\vspace{-0.25cm}

\subsection{Updating Solution Step} \label{Sec:OP_Overall_Algorithm}\vspace{-0.15cm}
The solution to $\mathcal{OP}$ for the set of variables $\vect{X}_q$, is computed for each algorithmic iteration $t+1$ and for each $q$ as 
follows:
\begin{equation} \label{eqn:OP1_ascent_solution}
	\vect{X}_q^{t+1} = \vect{X}_q^t + \alpha^t\left( \widehat{\vect{X}}_q^t - \vect{X}_q^t \right),
\end{equation}
where $\alpha^t$ represents the possibly time-varying step size.\vspace{-0.15cm}

\section{Numerical Results} \label{Sec:Numerical} \vspace{-0.15cm}
In our simulations, all nodes were considered positioned on a $3$D Cartesian coordinate system. In particular, we have set $Q \!=\! 4$ and located the BSs in a square of width $w \!=\! 60\,m$ placing BS$_1$ at the origin and the others at the remaining corners, letting $z_{\rm BS_q} \!=\! 5\,m$ $\forall q\!=\!1,2,\ldots,Q$. For simplicity, we considered $L_q \!=\! 1$ $\forall q=1,2,\ldots,Q$, and the UEs were located at the corners of a square, with origin at $(30,60)$ and width equal to $2.5\,m$, letting also $z_{\rm UE}=1.5\,m$. Each RIS was placed close to the corresponding BS with $z_{\rm RIS}=3\,m$: RIS$_1$ was fixed at $(-2.5,8.5)$, RIS$_2$ at $(62.5,8.5)$, RIS$_3$ at $(-2.5,111.5)$, and RIS$_4$ at $(62.5,111.5)$. All wireless wideband channels were modeled as described in \cite{katsanos2024multi} with $16$ delay taps. For the fading component, we have considered distance-dependent pathloss between any two nodes $i,\,j$ with distance $d_{i,j}$ (where $i,j \in \{\rm BS,UE,RIS\}$): ${\rm PL}_{i,j} \!=\! {\rm PL_0}(d_{i,j}/d_0)^{-\alpha_{i,j}}$ with ${\rm PL_0} \!=\! (\frac{\lambda_c}{4\pi})^2$ denoting the signal attenuation at the reference distance $d_0 \!=\! 1$ $m$ and $\lambda_c$ represents the carrier wavelength, with $f_c \!=\! 3.5$ GHz. For the pathloss exponents, we have set $\alpha_{\rm BS,UE} \!=\! 3.7$, $\alpha_{\rm BS,RIS} \!=\! 2.2$, and $\alpha_{\rm RIS,UE} \!=\! 2.6$. Equal transmit powers and noise variances was considered for all users: $P_q = P$ and $\sigma_{\ell_q,k}^2\!=\!-90$ dBm ($\forall$$k,\ell,q$), as well as bandwidth ${\rm BW}\!=\!0.1$ GHz and the number $K$ of SCs was set to $64$. For the algorithmic parameters, we have set $\tau\!=\!0.80$ and a time-varying step size (as detailed in \cite{scutari2013decomposition}). The RIS circuit elements were set as in \cite{abeywickrama2020intelligent}. For comparison purposes, we have also included the achievable rates for the following schemes: \textit{i}) ``w/o RISs'' with no RISs deployed; and \textit{ii}) ``RISs'' for $\vect{S}_q\!=\!\vect{I}_M$. We have also simulated the equivalent non-cooperative schemes for which ``$\vect{\Pi}\!=\!\vect{0}$''. We have used $100$ independent Monte Carlo realizations for all performance evaluation results that follow.
\begin{figure}[!t]
	\centering
	\includegraphics[width=3.20in]{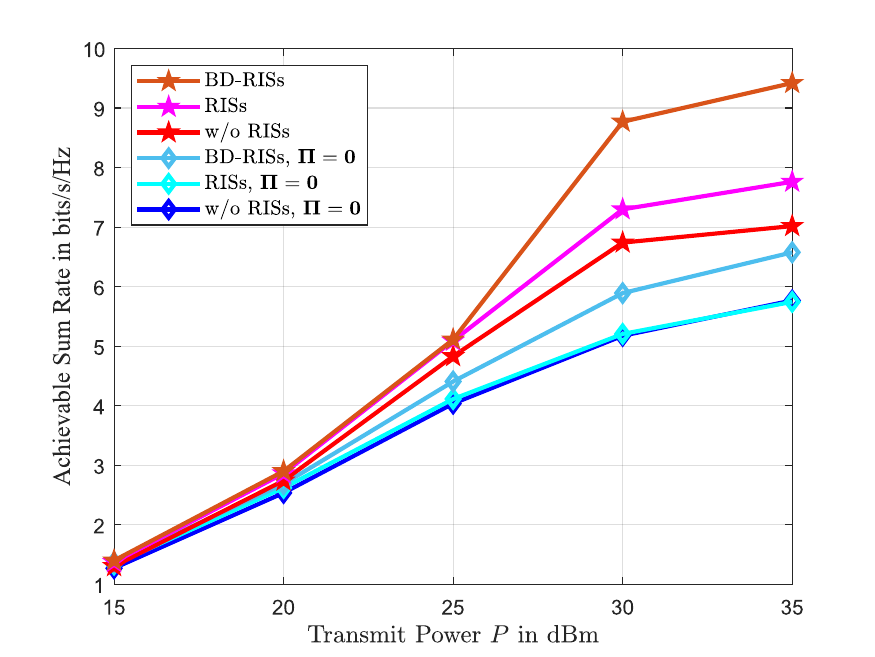}
	\caption{\small{Achievable sum-rate performance for $Q=4$ BSs and RISs, each with $N\!=\!4$ antennas and $M\!=\!100$ unit elements, respectively.}\vspace{-0.15cm}}
	\label{fig:Rates_vs_P}
\end{figure}

In Fig.~\ref{fig:Rates_vs_P}, we examine the performance of the proposed design as a function of each BS's transmit power $P$ for the various simulated cases. Evidently, all curves follow a non-decreasing trend as $P$ gets larger. It is also demonstrated that the achievable sum rate for the ``BD-RISs'' case outperforms the cases with diagonal RISs and that of ``w/o RISs,'' especially when $P\geq25$ dBm. This implies that the distributed schemes outperform the corresponding non-cooperative ones, indicating that adequately optimized cooperative transmit/reflective beamforming yields improved gains.

\section{Conclusion} \label{Sec:Conclusion} \vspace{-0.07cm}
In this paper, we studied the RIS-empowered interference broadcast channel and presented a cooperative approach for the achievable sum-rate maximization with wideband transmissions. Our numerical investigation showcased the additional degrees of freedom offered by the proposed optimized BD RISs in the high transmit power regime, as well as the gains offered by the cooperation among the multiple BSs.
\vspace{-0.05cm}

\bibliographystyle{IEEEtran}
\vspace{-0.1cm}
\bibliography{IEEEabrv,refs_SPAWC24}

\begin{thebibliography}{10}
\providecommand{\url}[1]{#1}
\csname url@samestyle\endcsname
\providecommand{\newblock}{\relax}
\providecommand{\bibinfo}[2]{#2}
\providecommand{\BIBentrySTDinterwordspacing}{\spaceskip=0pt\relax}
\providecommand{\BIBentryALTinterwordstretchfactor}{4}
\providecommand{\BIBentryALTinterwordspacing}{\spaceskip=\fontdimen2\font plus
\BIBentryALTinterwordstretchfactor\fontdimen3\font minus
  \fontdimen4\font\relax}
\providecommand{\BIBforeignlanguage}[2]{{%
\expandafter\ifx\csname l@#1\endcsname\relax
\typeout{** WARNING: IEEEtran.bst: No hyphenation pattern has been}%
\typeout{** loaded for the language `#1'. Using the pattern for}%
\typeout{** the default language instead.}%
\else
\language=\csname l@#1\endcsname
\fi
#2}}
\providecommand{\BIBdecl}{\relax}
\BIBdecl

\bibitem{CMY+24}
B.~Clerckx \emph{et~al.}, ``Multiple access techniques for intelligent and
  multi-functional 6{G}: Tutorial, survey, and outlook,'' \emph{arXiv
  preprint:2401.01433}, 2024.

\bibitem{RISoverview2023}
E.~Basar \emph{et~al.}, ``Reconfigurable intelligent surfaces for {6G}:
  Emerging applications and open challenges,'' \emph{IEEE Veh. Techcnol. Mag.},
  2024.

\bibitem{RIS_challenges}
G.~C. Alexandropoulos \emph{et~al.}, ``{RIS}-enabled smart wireless
  environments: Deployment scenarios, network architecture, bandwidth and area
  of influence,'' \emph{EURASIP J. Wireless Commun. and Netw.}, vol. 2023,
  no.~1, pp. 1--38, 2023.

\bibitem{LSN+23}
H.~Li \emph{et~al.}, ``Reconfigurable intelligent surfaces 2.0: Beyond diagonal
  phase shift matrices,'' \emph{IEEE Commun. Magazine}, 2023.

\bibitem{katsanos2022wideband}
K.~D. Katsanos \emph{et~al.}, ``Wideband multi-user {MIMO} communications with
  frequency selective {RIS}s: Element response modeling and sum-rate
  maximization,'' in \emph{Proc. IEEE ICC}, Seoul, South Korea, 2022.

\bibitem{DRM+24}
A.~S. {de Sena} \emph{et~al.}, ``Beyond diagonal {RIS} for multi-band
  multi-cell {MIMO} networks: A practical frequency-dependent model and
  performance analysis,'' \emph{arXiv preprint:2401.06475}, 2024.

\bibitem{li2024wideband}
H.~Li \emph{et~al.}, ``Wideband modeling and beamforming for beyond diagonal
  reconfigurable intelligent surfaces,'' \emph{arXiv preprint:2403.12893},
  2024.

\bibitem{li2022_nonDiag_switches}
Q.~Li \emph{et~al.}, ``Reconfigurable intelligent surfaces relying on
  non-diagonal phase shift matrices,'' \emph{IEEE Trans. Veh. Technol.},
  vol.~71, no.~6, pp. 6367--6383, 2022.

\bibitem{abeywickrama2020intelligent}
S.~Abeywickrama \emph{et~al.}, ``Intelligent reflecting surface: Practical
  phase shift model and beamforming optimization,'' \emph{IEEE Trans. Commun.},
  vol.~68, no.~9, pp. 5849--5863, 2020.

\bibitem{scutari2013decomposition}
G.~Scutari \emph{et~al.}, ``Decomposition by partial linearization: Parallel
  optimization of multi-agent systems,'' \emph{IEEE Trans. Signal Process.},
  vol.~62, no.~3, pp. 641--656, 2014.

\bibitem{tam2016_successive}
H.~H.~M. Tam \emph{et~al.}, ``Successive convex quadratic programming for
  quality-of-service management in full-duplex {MU}-{MIMO} multicell
  networks,'' \emph{IEEE Trans. Commun.}, vol.~64, no.~6, pp. 2340--2353, 2016.

\bibitem{PLS2022_counteracting}
G.~C. Alexandropoulos \emph{et~al.}, ``Counteracting eavesdropper attacks
  through reconfigurable intelligent surfaces: A new threat model and secrecy
  rate optimization,'' \emph{IEEE Open J. Commun. Soc.}, vol.~4, 2023.

\bibitem{katsanos2024multi}
K.~D. Katsanos, P.~Di~Lorenzo, and G.~C. Alexandropoulos,
  ``Multi-{RIS}-empowered multiple access: A distributed sum-rate maximization
  approach,'' \emph{arXiv preprint:2406.19334}, 2024.

\end{thebibliography}

\end{document}